\documentstyle[aps,prb,twocolumn,graphics]{revtex}

\begin{document}

\twocolumn[\hsize\textwidth\columnwidth\hsize\csname@twocolumnfalse\endcsname 

\title{\ Nonperiodic Flux to Voltage Conversion of an Arithmetic Series Array of dc SQUIDs}

\author{Ch. H\"{a}ussler, J. Oppenl\"{a}nder\thanks{Electronic mail: joerg.oppenlaender@uni-tuebingen.de}, and N. Schopohl}

\address{Institut f\"{u}r Theoretische Physik, Eberhard-Karls-Universit\"{a}t T\"{u}bingen,
Auf der Morgenstelle 14, 72076 T\"{u}bingen, Germany}

\maketitle

\begin{abstract}
A theoretical study on the voltage response function \( \left\langle V\right\rangle  \)
of a series array of dc SQUIDs is presented in which the elementary dc SQUID loops
vary in size and, possibly, in orientation. Such series arrays of
two-junction SQUIDs possess voltage response functions vs. external magnetic
field \( {\bf B} \) that differ substantially from those of corresponding regular
series arrays with identical loop-areas, while maintaining a large voltage swing
as well as a low noise level. Applications include the design of current amplifiers
and \emph{quantum interference filters}.
\bigskip
\end{abstract}
]

Recently series arrays of dc SQUIDs have been successfully exploited as current 
amplifiers with wide bandwidth, large dynamic range and low noise level \cite{hirayama}. 
Using thin-film Nb-technology, amplifiers consisting out of up to \( 10^{3} \)
identical dc SQUID loops have been fabricated. Such serial devices are 
characterized by large voltage swings of several \( mV \) and current-to-voltage 
transfer functions of some \( V/mA \) so that a direct connection to a room 
temperature preamplifier is feasible. 

At present only series arrays consisting out of \emph{identical} dc SQUID loops have been described
in the literature \cite{hirayama}. The dc voltage response function 
\( \left\langle V\right\rangle  \) of such regular series arrays displays a 
\( \Phi _{0} \)-periodicity just as a single dc SQUID where \( \Phi _{0}=\frac{h}{2e} \) 
is the elementary flux quantum \cite{clarke}.
However, this periodicity may represent a serious limitation for the modes of operation of such devices.
To get a linear flux-to-voltage conversion a feedback 
circuit is used in most applications in which the dc SQUID acts as a null-flux detector. 
Furthermore special electronic devices and efficient
background shielding is often required \cite{clarke}. Here theoretical studies on the voltage response 
of arithmetic and irregular series arrays of dc SQUID loops are presented, in which the individual 
loop-areas are \emph{not} all equal. For such devices the advantages of dc SQUID series arrays are
preserved but limitations due to the periodicity of the voltage response can be circumvented. 

\begin{figure}
{\par\centering \resizebox*{0.47\textwidth}{!}{\includegraphics{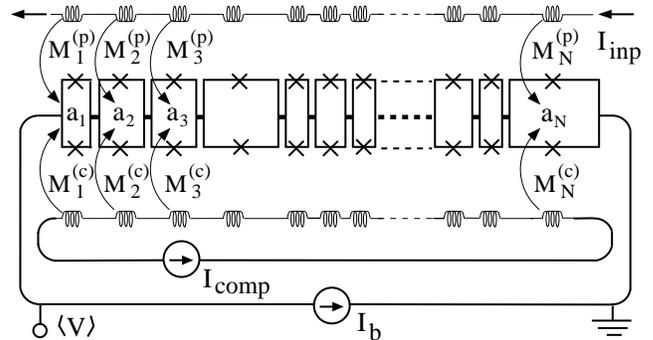}} \par}
\caption{\label{figure1}Schematic diagram of a generic planar series array of two-junction
SQUIDs. \( {\bf a}_{n} \) denotes the orientated area element of the \( n^{th} \) dc SQUID loop.
The Josephson junctions are indicated by crosses (x). The array is biased
with the current \protect\( I_{b}\protect \) and produces an dc output voltage
\protect\( \left\langle V\right\rangle \protect \). The magnetic flux is coupled into 
the dc SQUID loops by \protect\( I_{inp}\protect \) flowing in a common input
flux bias line and also by \protect\( I_{comp}\protect \) in a control
flux bias line. }
\end{figure}

The arrays under consideration consist of \( N \) two junction SQUID loops
connected in series. The bias current \( I_{b} \) is fed into the array as
indicated in Fig.(\ref{figure1}). In general the areas of the \( N \) loops in a \emph{generic} array differ
in size and, possibly, in orientation. Let \( {\bf a}_{n} \) be the orientated
area element of the \( n^{th} \) dc SQUID loop. The magnetic flux threading \( {\bf a}_{n} \) 
is then \( \Phi _{n}=\left\langle {\bf B},\, {\bf a}_{n}\right\rangle  \),
where \( {\bf B} \) is the total magnetic field. Taking into account all inductive
effects in the dc SQUID array the total magnetic field 
\( {\bf B}={\bf B}^{(p)}+{\bf B}^{(s)}+{\bf B}^{(c)} \)
is a superposition of the primary magnetic field \( {\bf B}^{(p)} \) one wants
to measure, the secondary magnetic field \( {\bf B}^{(s)} \) induced by currents
that flow in the dc SQUID array and, possibly, the magnetic compensation field
\( {\bf B}^{(c)} \) induced by compensation current(s) \( I_{comp} \)
flowing through a set of suitable orientated compensation coils or wires. 

As a special case a generic \emph{planar} dc SQUID array is shown in Fig.(\ref{figure1}). 
The primary magnetic flux is coupled into the individual dc SQUID loops by the
signal or input current \( I_{inp} \) which flows through a common input flux
bias line. In typical applications \( I_{inp} \) is the current provided by a pick-up loop or
by other signal sources. In addition the flux in the individual dc SQUID loops may
be modulated by the current \( I_{comp} \) flowing through a common compensation
or control flux bias line. 

Let \( M^{(c)}_{n} \) and \( M_{n}^{(p)} \) denote
mutual inductances of the currents \( I_{comp} \) and \( I_{inp} \) associated with
the \(n^{th} \) loop with area-element \( {\bf a}_{n} \) and let \( {\bf L}_{A} \) be the inductance
matrix describing all other inductive couplings in the circuit \cite{oppenlaender}. 
The total magnetic flux in the \( n^{th} \) loop is then given by 
\( \Phi _{n}=M^{(p)}_{n}\, I_{inp}+M^{(c)}_{n}\, I_{comp}+{\bf L}_{A}\, {\bf I} \),
where \( {\bf I}=(I_{1},\ldots ,I_{2N}) \) is the vector of the \( 2N \) currents
\( I_{n} \) through the Josephson junctions. 
Since \( I_{comp} \) should be able to compensate the primary induced magnetic fluxes
\( M_{n}^{(p)}\, I_{inp} \) in all loops simultaneously the relation 
\( M_{n}^{(c)}=\alpha \, M_{n}^{(p)} \)
should hold for all \( n \), where \( \alpha  \) is a constant. 

Within the range of validity of the RCSJ-model \cite{likharev} the current
\( I_{m} \) through the \( m^{th} \) array junction with gauge invariant phase difference
\( \varphi _{m} \), critical current \( I_{c,m} \), normal resistance \( R_{m} \), and
 capacitance \( C_{m} \) is given by
\begin{equation}
\label{eq0}
I_{m}=I_{c,m}\, \sin (\varphi _{m})+
\frac{\hbar }{2eR_{m}}\, \partial _{t}\varphi _{m}+
\frac{\hbar C_{m}}{2e}\, \partial _{t}^{2}\varphi _{m}.
\end{equation}
For simplicity identical parameters for the junctions are supposed in the following discussion:
\( I_{c,m}=I_{c} \), \( R_{m}=R \) and \( C_{m}=C \).
The dimensionless parameter 
\( \beta _{C}=\frac{2\pi I_{c}R^{2}C}{\Phi _{0}} \)
characterizes then the capacitance effect of the junctions within this model \cite{likharev}.

In the overdamped regime, \( C=0 \), and for vanishing inductive coupling in
the array, \( {\bf L}_{A}=0 \), an analytical expression for the dc voltage response
function \( \left\langle V\right\rangle  \) of the dc SQUID series can be determined. 
Under conditions where a constant current \( I_{b}>2\, I_{c} \) is biased, 
and assuming for simplicity a static magnetic field 
\( {\bf B}={\bf B}^{(p)}+{\bf B}^{(c)} \), the output voltage
\( \left\langle V\right\rangle _{n} \) of the \( n^{th} \) dc SQUID in the array
as a function of the bias current \( I_{b} \) and of the applied flux 
\( \Phi _{n} \) is given by \cite{dewaele}
\begin{equation}
\label{eq1}
\left\langle V\right\rangle _{n}=
I_{c}\, R\, \sqrt{\left( \frac{I_{b}}{2\, I_{c}}\right) ^{2}-
\left| \cos (\pi \, \frac{\Phi _{n}}{\Phi _{0}})\right| ^{2}}.
\end{equation}
The voltage response \( \left\langle V\right\rangle  \) of a series array
containing \( N \) dc SQUIDs is then given as the sum of the voltages
\( \left\langle V\right\rangle _{n} \):
\begin{equation}
\label{eq2}
\left\langle V\right\rangle =\sum _{n=1}^{N}\left\langle V\right\rangle _{n}.
\end{equation}
Since the maximum voltage swing increases linearly with the number \( N \) of dc SQUIDs
in the array the gain 
\( \frac{\partial \left\langle V\right\rangle }{\partial I_{inp}} \)
is proportional to \( N \). However, the noise in the array increases only as 
\( N^{\frac{1}{2}} \),
because the noise voltages implied by the junctions in the individual SQUID-loops of 
the array are expected to add incoherently \cite{hirayama}. 
These features make generic series arrays attractive for amplifiers and magnetometers.

Consider, as a special case, a regular array, consisting out of \( N \) \emph{identical}
dc SQUID loops so that \( {\bf a}_{n}={\bf a}_{N} \) with identical mutual inductances
\( M_{n}^{(c)}=M_{N}^{(c)}\) and \(M_{n}^{(p)}= M_{N}^{(p)} \) of the flux bias lines.
As a result the flux threading each individual loop is then given by \( \Phi _{n}=\Phi _{N} \).
In this case the voltage response, \( \left\langle V\right\rangle =N\, \left\langle V\right\rangle _{N} \),
displays as a function of flux a \( \Phi _{0} \)-periodicity.

A voltage response function \( \left\langle V\right\rangle \) 
with a much longer period may be obtained in an
\emph{arithmetic} dc SQUID series array where the orientated area elements increase
in size according to the arithmetic relation
\begin{equation}
\label{eq3}
{\bf a}_{n}=\frac{n}{N}\, {\bf a}_{N}, \;\;\; n=1,\ldots,N.
\end{equation}
Provided the mutual inductances are chosen such that 
\( M_{n}^{(c)}=\frac{\left| {\bf a}_{n}\right| }{\left| {\bf a}_{N}\right| }\, M_{N}^{(c)} \) and
\( M_{n}^{(p)}=\frac{\left| {\bf a}_{n}\right| }{\left| {\bf a}_{N}\right| }\, M_{N}^{(p)} \)
the fluxes are determined by 
\( \Phi _{n}=\frac{n}{N}\, \Phi _{N} \),
where \( \Phi _{N} \) is the flux through the largest area element \( {\bf a}_{N} \).
In this case the individual dc SQUIDs operate with different periodicity in flux.
This implies that the voltage response \( \left\langle V\right\rangle  \) of the whole
arithmetic array as a function of \( \Phi _{N} \) is periodic 
with period \( N\, \Phi _{0} \).
This period is \( N \) times larger than for a regular array, consisting
of \( N \) identical dc SQUID loops with size \( {\bf a}_{N} \). 

\begin{figure}
{\par\centering \resizebox*{0.47\textwidth}{!}{\includegraphics{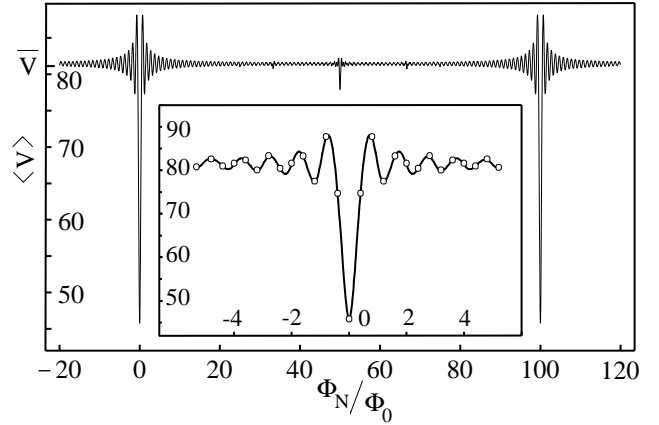}} \par}
\caption{\label{figure2}Voltage response \protect\( \left\langle V\right\rangle \protect \)
according to Eq.(\ref{eq2}) in units of \protect\( I_{c}\, R\protect \) vs. external flux \protect\( \Phi _{N}\protect \)
through the largest area element \protect\( {\bf a}_{N}\protect \) for an arithmetic
series array with \protect\( N=100\protect \) dc SQUIDs  for bias current \protect\( I_{b}=2.2\, I_{c}\protect \)
and vanishing inductive coupling \protect\( (\beta _{L,max}=0\protect) \). The
dc SQUID loops increase in size according to the linear relation 
\protect\( {\bf a}_{n}=\frac{n}{N}\, {\bf a}_{N}\protect \).
In the inset the theoretical curve (solid line, Eq.(\ref{eq2})) is compared with the analytical
approximation (circles, Eq.(\ref{eq4})).}
\end{figure}

In Fig.(\ref{figure2}) the voltage response for an arithmetic series array with
\( N=100 \) dc SQUIDs is shown. The bias current \(I_b\) was adjusted for maximum 
voltage swing slightly above \( 2\, I_{c} \). In the depicted interval the 
voltage response shows pronounced antipeaks at 
\( \Phi _{N}=0 \) and at \( \Phi _{N}=N\, \Phi _{0} \).
Between these two peaks the output voltage oscillates with a very small amplitude
around an average voltage drop \( \overline{V} \).  
For a sufficiently high number of elements in the arithmetic array the flux-distribution
over the dc SQUID loops is uniform between some \( \Phi _{min} \) and \( \Phi _{max} \).
Let \( \kappa =\frac{\Phi _{min}}{\Phi _{max}} \) be the flux spread coefficient,
\( f=\pi \, \frac{\Phi _{max}}{\Phi _{0}} \) the frustration of the greatest
loop in the array and let \( \omega _{0}=\sqrt{\left( \frac{I_{b}}{2\, I_{c}}\right) ^{2}-1} \)
be the characteristic frequency of a single dc SQUID for zero flux. If the frustration
\( f \) is limited to the periodicity interval \( \left| f\right| \leq \pi \, \frac{N}{2} \) the
voltage response \( \left\langle V\right\rangle  \) as a function of \( f \)
can be approximated by an integral:
\begin{eqnarray}
\label{eq4}
\left\langle V\right\rangle &\approx& \frac{N\, I_{c}\, R}{\Phi _{max}-\Phi _{min}}
\int\limits ^{\Phi _{max}}_{\Phi _{min}}\sqrt{\left( \frac{I_{b}}{2I_{c}}\right) ^{2}-
\left| \cos (\pi \frac{\Phi }{\Phi _{0}})\right| ^{2}}d\Phi \nonumber \\
 &=& N\, I_{c}\, R\, \frac{\omega _{0}}{(1-\kappa )}\frac{1}{f}
\left[ E\left( f,m\right) -E\left( \kappa f,m\right) \right] ,
\end{eqnarray}
where \(m=-\omega _{0}^{-2}\) and \( E(f,m) \) denotes the elliptic integral of the second kind \cite{abram}. 
The voltage versus flux relation displayed in Fig.(\ref{figure2}) shows an excellent agreement between 
the theoretical result Eq.(\ref{eq2}) and its analytical approximation Eq.(\ref{eq4}). For
bias currents \( I_{b}=2.2\, I_{c} \) there results an average voltage drop of 
\( \overline{V}=0.81\, I_{c}\, R\, N \).

At first sight all the above results apply only in the overdamped junction regime and only
for vanishing inductive couplings in the arithmetic dc SQUID array, i.e. for \( \beta _{C}=0 \)
and \( {\bf L}_{A}=0 \). A dimensionless measure for the inductive effects
in a single dc SQUID loop with loop-inductance \( L \) is \( \beta _{L}=\frac{L\, I_{c}}{\Phi_{0}} \).
Dependent on \( \beta _{L} \) there exists an optimal size for any dc SQUID loop
\cite{clarke} which should coincide with the maximal loop size \( {\bf a}_{max} \)
in a series array. The corresponding maximal \( \beta _{L,max} \) is then a
measure of the self- and mutual-inductive couplings among the currents flowing
in the dc SQUID array.

Taking into account all inductive couplings \( {\bf L}_{A} \) in the arithmetic
array our computer simulations of the full non-linear dynamics \cite{oppenlaender} of the
\( 2\, N \) coupled Josephson phases \( \varphi _{i} \) reveal that expression
Eq.(\ref{eq4}) compares qualitatively well with the calculated voltage response
functions. In particular the long periodicity of
the voltage \( \left\langle V\right\rangle  \) vs. flux \( \Phi _{N} \) relation
is not affected by inductive effects. Qualitatively similar behavior was found for a parallel
multi-junction interferometer \cite{haeussler}. 

\begin{figure}[t]
{\par\centering \resizebox*{0.47\textwidth}{!}{\includegraphics{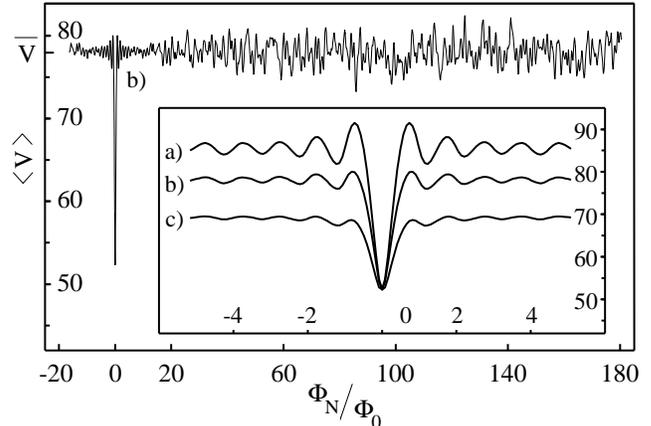}} \par}
\caption{\label{figure3}Voltage response \protect\( \left\langle V\right\rangle \protect \)
in units of \protect\( I_{c}\, R\protect \) vs. external flux \protect\( \Phi _{N}\protect \)
for an arithmetic series array with \protect\( N=100\protect \) for \protect\( I_{b}=2.2\, I_{c}\protect \),
\protect\( \beta _{C}=0.5\protect \), and \protect\( \beta _{L,max}=0.3\protect \).
Small random fluctuations were added to the loop size distribution \({\bf a}_{n}=\frac{n}{N}\, {\bf a}_{N}\).
The inset depicts \protect\( \left\langle V\right\rangle \protect \) for
various inductive couplings around the global minimum at \(\Phi_N=0\): a) \protect\( \beta _{L,max}=0\protect \),
b) \protect\( \beta _{L,max}=0.3\protect \), and c) \protect\( \beta _{L,max}=0.7\protect \).}
\end{figure}

As far as the irregularity is concerned, we also find that
in an arithmetic array  \( \left\langle V\right\rangle  \) is very responsive to adding 
small random fluctuations to the size distributions of the area elements. In this case 
\( \left\langle V\right\rangle  \)
becomes \emph{nonperiodic} with a pronounced antipeak only around \( \Phi_N=0 \).
For fixed \( \beta _{C} \), but different strengths of the magnetic coupling
\( \beta _{L,max} \), Fig.(\ref{figure3}) shows the voltage response for an
arithmetic planar series array with \( N=100 \) dc SQUIDs where small random fluctuations
were added to the loop size distribution Eq.(\ref{eq3}). 

For fixed \( \beta _{C} \) the global minimum of \( \left\langle V\right\rangle\) at \( \Phi _{N}=0 \)
depends only slightly on the strength \( \beta _{L,max} \) of the magnetic
coupling whereas the voltage branch \( \left\langle V\right\rangle \approx \overline{V} \)
depends strongly on \( \beta _{L,max} \),
as can be seen in the inset of Fig.(\ref{figure3}). For increasing \( \beta _{L,max} \)
the voltage swing and the maximum of the voltage transfer function 
\( \frac{\partial \left\langle V\right\rangle }{\partial \Phi _{N}} \)
decrease as it is the case
for conventional single dc SQUIDs when \( \beta _{L} \) increases \cite{enpuku}. If
on the other hand \( \beta _{L,max} \) is kept constant the global minimum of
\( \left\langle V\right\rangle\) at \( \Phi _{N}=0 \) is an increasing function of
\( \beta _{C} \) as a comparison of Fig.(\ref{figure2}) and Fig.(\ref{figure3})
reveals. In addition Fig.(\ref{figure3}a) shows an enhancement of the branch
\( \left\langle V\right\rangle \approx \overline{V} \) for \( \beta _{C}=0.5 \)
in comparison to the underdamped case, since for single dc SQUIDs the effect of
the capacitance \( \beta _{C} \) is to increase the voltage near integer values
of the applied flux \cite{enpuku}. Although the total area
\(A_{tot}=\frac{N+1}{2} \, \left| {\bf a}_N \right|\) of an arithmetic array 
is smaller by a factor of \(2\) compared to a regular array with 
\(A_{tot}=N \, \left| {\bf a}_N \right |\), this effect provides that the voltage transfer function 
of both arrays is comparable. 
This suggests a higher integration density for arithmetic array circuits on chip.

In summary, arithmetic series arrays of two-junction SQUIDs possess voltage response functions
with a much longer period with respect to the applied magnetic field than regular series
arrays, while maintaining a comparable transfer function and a low noise level.
In particular if the loop-sizes or loop-orientations are distributed
randomly the voltage response becomes nonperiodic with a single pronounced antipeak
only around zero applied magnetic field. These features are preserved when all capacitive and inductive effects are taken into account. 
Therefore arithmetic or irregular series arrays of dc SQUIDs
can be used as \emph{quantum interference filters} for various applications, including, e.g., the
relatively simple and extremely sensitive measurement of the \emph{absolute} strength of magnetic fields.
 
\textbf{Acknowledgments}: We thank R. P. Huebener, R. Kleiner and T. Tr\"{a}uble
for useful discussions. Support by ''Forschungsschwerpunktprogramm des Landes
Baden-W\"{u}rttemberg'' is gratefully acknowledged.


\begin{thebibliography}{100}
\bibitem[1]{hirayama}F. Hirayama, N. Kasai, M. Koyanagi, IEEE Trans. Appl. Supercon. \textbf{9},
2923, (1999); R. P. Welty, J. M. Martinis, IEEE Tans. Appl. Supercon. \textbf{3}, 2605, (1993);
R. P. Welty, J. M. Martinis, IEEE Tans. Magn. \textbf{27}, 2924, (1991). 
\bibitem[2]{clarke}J. Clarke, in 'The New Superconducting Electronics', edited by H. Weinstock
and R. W. Ralston, Kluwer Academic Publishers, (1993).
\bibitem[3]{oppenlaender}J. Oppenl\"{a}nder, Ch. H\"{a}ussler, N. Schopohl, J. Appl. Phys. \textbf{86}, 5775
(1999).
\bibitem[4]{likharev}K. K. Likharev, in 'Dynamics of Josephson Junctions and Circuits', Gordon and
Breach Science Publishers, 2nd printing (1991).
\bibitem[5]{dewaele}A. Th. A. M. De Waele, R. De Bruyn Ouboter, Physica \textbf{41}, 225 (1969).
\bibitem[6]{abram} M. Abramowitz, A. Stegun (eds.), in 'Handbook of Mathematical Functions', 
Applied mathematics series of the National Bureau of Standards (1972).
\bibitem[7]{haeussler} Ch. H\"{a}ussler, J. Oppenl\"{a}nder, and N. Schopohl, cond-mat/0003487, (2000).
\bibitem[8]{enpuku}K. Enpuku, K. Sueoka, K. Yoshida, and F. Irie, J. Appl. Phys. \textbf{57},1691
(1985).
\end{thebibliography}
\end{document}